# MARS Star Density Results for Shielding Applications

**Diane Reitzner**
**7/28/2010**

Comparison of MARS 15.09 star density results in thick shielding to the star density distribution in iron and concrete rule-of-thumb expectations.

# Introduction

MARS' validity with respect to shielding applications had been assessed years ago for studies of the NuMI beam line.  In the intervening years, MARS has undergone several updates.  The current release is 15.07 and the beta version 15.09 will be publically released in the very near future   Given the number of iterative releases of MARS in the intervening years, it may be prudent to investigate whether the current MARS version is valid.

Past experiments have provided a general "rule-of-thumb" (RoT) for the star density distributions in iron and concrete and may be used to test the validity of the current MARS code.  For concrete, the RoT calls for the star density to decrease by a factor of 10 for every 91.44 cm (1 yard) of concrete.  For iron, a similar decrease is expected to be seen for every 30.48 cm (1 foot).  Note that these RoT are for secondary particles.  MARS version 15.09 is used to test the star density distribution because it will be replacing the current release 15.07 in the very near future.  Furthermore a coding error in MARS 15.07 selects an inappropriate physics models when the thick shielding option is turned on.

120 GeV protons were generated in a simple geometry model of the LNBE decay pipe.  The model includes a 0.97 m long carbon target of radius 7.5 mm, and two NuMI like focusing horns upstream of a 250 m air filled pipe.   The target and horns are used to generate a proton beam envelope that mimics realistic beam conditions.   The pipe has a 2 m radius and is surrounded by 2 m of shielding.  The shielding is subdivided into 5 cm thick rings.  The thick shielding option (IND=6) is set to true in the input file MARS.INP.  The average star density ($S_{av}$) was extracted for each subdivision of the shielding and fit to the function:

$$\log_{10} S_{av} = m(r-300) + b$$

**Equation 1**

where $b$ and $m$ are the fitted parameters, and $r$ is the radial position of the shielding subdivision with respect to the centre of the pipe.  The offset of 300 cm reduces the lever arm of the fit thus reducing the uncertainties on the fitted parameters.  The amount of shielding which reduces $S_{av}$ by a factor of 10 is equal to 1/$m$.

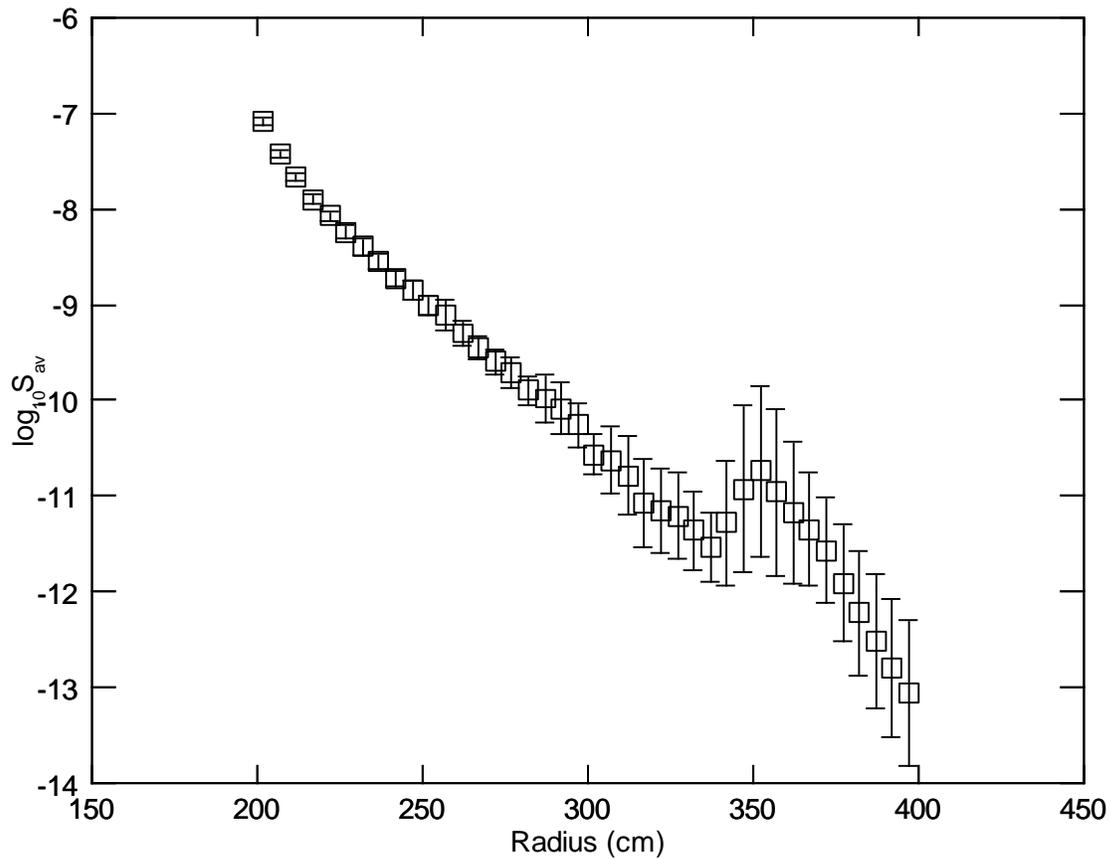

**Figure 1** $\log_{10}(S_{av})$ vs R for iron

# Iron

$10^4$ protons were generated for the iron shielding study. The shielding starts at a radius of 200 cm and ends at 400 cm. The fit to equation 1 was preformed over the region 220 < r < 300 cm. Figure 1 shows the results for $\log_{10}S_{av}$ vs $r$. The "peak" at 350 cm is an artifact of the MARS star density estimation routines used in regions of low statistics and can be removed by increasing the statistics. Because of this, summation of multiple low statistics runs should be avoided. The Increased slope at the air-iron interface is due to build up from high energy charged particles that dominate the radiation in shallow layers of shielding. The slope *m* is found to be $(-2.979\pm0.033)10^{-2}$ cm$^{-1}$ which gives the thickness to reduce $S_{av}$ by a factor of 10 to be 33.6 ±0.4 cm.

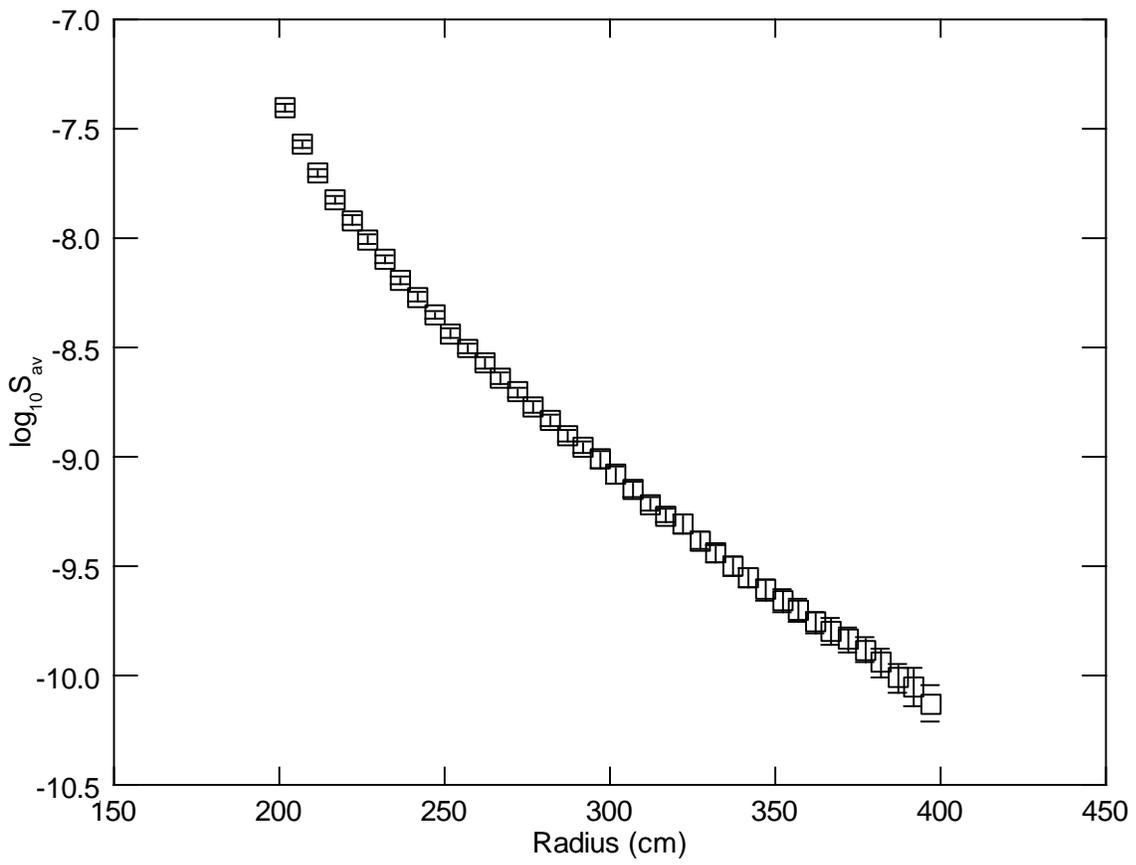

Figure 2 Log$_{10}$(S$_{av}$) vs R for concrete

## Concrete

For concrete shielding of density 2.35 g/cm³, 5×10⁴ events were generated and the fit to equation 1 was performed for r > 250 cm. Like the case for iron shielding, the concrete shielding is located from 200 to 400 cm from the centre of the pipe. Figure 2 shows the results for $\log_{10}S_{av}$ vs $r$. The higher statistics has eliminated the peak structure seen at large radii in the results for iron. The result of $(-1.148\pm0.013)\times10^{-2}$ cm$^{-1}$ for $m$ equates to requiring 87.1±1.0 cm of concrete to reduce $S_{av}$ by a factor of 10.

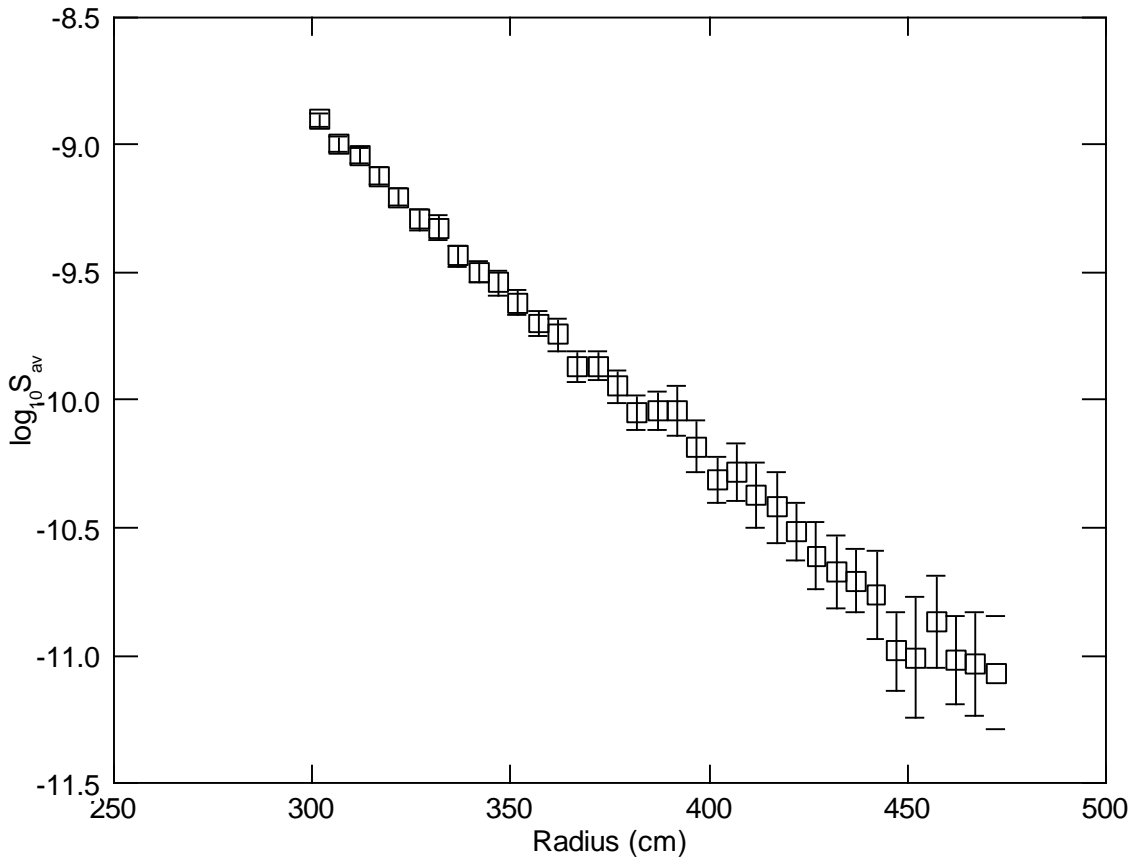

Figure 3 Log$_{10}$(S$_{av}$) vs R for dolomite

## Dolomite

This test was repeated for dolomite located behind 1 m of concrete shielding to confirm that the build-up seen in the concrete and iron shielding for $r$ < 220 cm is due to the radiation field being dominated by high energy charged particles in shallow layers of shielding. The shielding starts at a radius of 200 cm. 45×10³ events were generated and the results are shown in figure 3. Fitting of equation 1 gives a value of *m* to be $(-2.979\pm0.033)\times10^{-2}$ which corresponds to 70.8 ±1.2 cm of dolomite required to decrease on $S_{av}$ by a factor of 10.   Dolomite, being denser than concrete (2.85 g/cm$^3$ compared to 2.35 g/cm$^3$), reduces the radiation field by a greater amount per unit shield thickness.

## Conclusion

The results for the reduction of $S_{av}$ by a factor of 10 from the MARS MC are comparable to the RoT estimates.  The RoT guide lines are 91.44 cm (1 yard) of concrete, and 30.48 cm (1 foot) for iron.   The results from MARS are 87.1±1.0 cm and 33.6 ±0.4 cm respectively.   Estimation of the star densities by MARS in regions of low statistics with the thick shielding estimation turned on introduces anomalies in the star density distribution. To avoid questionable results, high statistics must be achieved in the regions of interest.  Achieving high statistics should not be done by summing a series of low statistics runs.  An uncertainty in the star density of 20% or more indicates insufficient statistics for summation.